\documentclass[twocolumn,showpacs,preprintnumbers,superscriptaddress]{revtex4}
\usepackage{graphicx}
\usepackage{dcolumn}
\usepackage{bm}
\usepackage{color}
\usepackage{amsmath}
\usepackage{amssymb}
\usepackage{amsfonts}
\usepackage{esint}
\usepackage{times}

\begin{document}
\title{Circuit QED with qutrit: coupling three or more atoms via virtual
photon exchange}
\author{Peng Zhao}
\affiliation{National Laboratory of Solid State Microstructures, \\
School of Physics, Nanjing University, Nanjing 210093, China}
\author{Xinsheng Tan}
\affiliation{National Laboratory of Solid State Microstructures, \\
School of Physics, Nanjing University, Nanjing 210093, China}
\author{Haifeng Yu}
\email{hfyu@nju.edu.cn}
\affiliation{National Laboratory of Solid State Microstructures, \\
School of Physics, Nanjing University, Nanjing 210093, China}
\affiliation{Synergetic Innovation Center of Quantum Information $\&$ Quantum Physics, \\
University of Science and Technology of China, Hefei, Anhui 230026, China}
\author{Shi-Liang Zhu}
\affiliation{National Laboratory of Solid State Microstructures, \\
School of Physics, Nanjing University, Nanjing 210093, China}
\affiliation{Synergetic Innovation Center of Quantum Information $\&$ Quantum Physics, \\
University of Science and Technology of China, Hefei, Anhui 230026, China}
\author{Yang Yu}
\affiliation{National Laboratory of Solid State Microstructures, \\
School of Physics, Nanjing University, Nanjing 210093, China}
\affiliation{Synergetic Innovation Center of Quantum Information $\&$ Quantum Physics, \\
University of Science and Technology of China, Hefei, Anhui 230026, China}
\date{\today}

\begin{abstract}
We present a model to describe a generic circuit QED system which consists
of multiple artificial three-level atoms, namely qutrits, strongly
coupled to a cavity mode. When the state transition of the atoms disobey the
selection rules the process that does not conserve the number of excitations
can happen determinatively. Therefore, we can realize coherent exchange
interaction among three or more atoms mediated by the exchange of virtual photons. In
addition, we generalize the one cavity mode mediated interactions
to the multi-cavity situation, providing a method to entangle atoms located
in different cavities. Using experimental feasible parameters, we
investigate the dynamics of the model including three cyclic-transition
three-level atoms, for which the two lowest-energy
levels can be treated as qubits. Hence, we have found that two qubits can jointly exchange
excitation with one qubit in a coherent and reversible way. In the whole
process, the population in the third level of atoms is negligible and the
cavity photon number is far smaller than 1. Our model provides a
feasible scheme to couple multiple distant atoms together, which may find
applications in quantum information processing.
\end{abstract}

\maketitle

\section{Introduction}

Light-matter interaction has been an active field for more than one century
\cite{R1}. Recent practice for realizing quantum information and quantum
computation attracts more attention on the atom-cavity systems \cite{R2,R3}.
The circuit quantum electrodynamics (QED) \cite{R4,R5}, where
superconducting qubits act as artificial atoms strongly coupled to
transmission line resonator, has been demonstrated as a well-controlled
atom-cavity system for studying quantum optics and quantum information
processing \cite{R6,R7}. By using circuit QED, people have engineered the
qubit-cavity coupling to control quantum state of the system, demonstrating
a series of remarkable achievements \cite{R7}. In addition, with the exchange of
virtual photons, two distant qubits, which are coupled to a common cavity
bus, can exist a strong flip-flop interaction \cite{R8,R9,R10,R11}. Based on
this photon-mediated interaction, people have realized two-qubit gate
and created two-qubit entanglement \cite{R12,R13}. Recently, more
complicated processes such as creating multi-qubit entanglement and
realizing multi-qubit gates, which include sequence of two-qubit gates along
with single qubit operations, have also been demonstrated in circuit QED
systems \cite{R14,R15}. With the increase of the complexity of the
operations, the decoherence of the system protrudes out as one of the big
challenges \cite{R16}.

In general, the quantum processors contain many qubits coupled through a cavity
or many cavities. A whole process involving multiple qubits that can quickly
couple multiple distant qubits together would be very promising to combat
the decoherence and reduce the number of gates in quantum algorithms
\cite{R17,R18,R19}. However, the conventional photon-mediated interaction
picture does not support this whole operation \cite{R20,R21,R22,R23}. In the
ordinary circuit QED model, superconducting qubits act as
two-level systems coupled to a cavity mode \cite{R4,R5}. The qubit-cavity
system is restricted to the situation where the qubit-cavity coupling
strength is much smaller than the qubit transition frequency and cavity
resonance frequency \cite{R4,R5}. In this coupling regime, it is usually
valid to apply the rotating-wave approximation (RWA) \cite{R24,R25}. For
system consisting of multi-qubit strongly coupled to a cavity mode, the physics
can be well described by the Tavis-Cummings Hamiltonian \cite{R25} under
the RWA. When we apply the RWA, counterrotating terms (i.e., the excitation number
nonconserving process and virtual transitions) have been dropped, and the
Hamiltonian conserves the total excitation number. Therefore, the
photon-mediated multi-qubit interactions can happen only if the process
conserves the excitation number, e.g., the two-qubit flip-flop interaction
\cite{R4} or the four-spin ring exchange interaction \cite{R26}. Multi-qubit
interactions which do not conserve the excitation number are prohibited.

Recently, Stassi \textit{et al.}$\,$\cite{R27} have theoretically
demonstrated that the interaction of multiple spatially-separated atoms can
be realized via the exchange of virtual photons in the ultrastrong-coupling
(USC) regime of cavity quantum electrodynamics \cite{R28}. In the USC regime,
the atom-cavity coupling strength is comparable to the atom and cavity
energy scales. Therefore, the usual RWA is no longer valid, and the
counterrotating terms have become relevant. The coherent exchange
interaction between multiple distant atoms via intermediate virtual states
connected by the counterrotating interaction terms, can happen deterministically.
Furthermore, the excitation number nonconserving process in the USC regime,
including multi-photon Rabi oscillation \cite{R29,R30}, a single photon
exciting multi-atom simultaneously \cite{R31}, quantum nonlinear optics with
atoms and virtual photons \cite{R32,R33} have also been theoretically predicted.
However, although a few experiments have recently achieved the USC regime in
solid-state quantum system \cite{R34,R35,R36}, quantum state manipulating
and high-fidelity readout are still a tough challenge with existing
technique, hindering the practical implementation of these coherent
exchange interactions at present.

In this paper, we theoretically demonstrate that the generalized multiple
distant atoms coupling via the exchange of virtual photons can be realized
in the conventional strong coupling regime of circuit QED. We consider a
generic circuit QED system which consists of multiple cyclic artificial three-level
atoms (qutrit), strongly coupled to a cavity mode. Having derived the
effective interaction Hamiltonian, we show that multiple atoms can conspire
to jointly exchange excitation with just one single atom, which obviously
does not conserve the total excitation number. The occurance of this process
indicates a resonant interaction among multiple atoms via the exchange of
virtual photons. The physics behind this virtual photon mediated interaction
is that the auxiliary third level of the cyclic qutrits allows a large number of virtual
transitions which do not conserve the excitation number contribute to the
effective coupling \cite{R37}. Furthermore, we have done numerical
simulation with experimentally feasible parameters in circuit QED systems.
It is found that the multiple atoms can jointly exchange
excitation with one atom with a probability approaching one. In the whole
process, the occupation of the third level of atoms and the cavity mode is
far smaller than 1. In addition, we generalize the above model to multi-cavity
case and find that the multi-cavity mediated interactions among three or more
atoms can also be realized in the strong coupling regime.

The rest of this paper is organized as follows. In Sec.$\,$II, we introduce
our model and derive the general form of the effective Hamiltonian. In Sec.$
\,$III, we apply the model to three-atom situation, where the
photon-mediated three-atom interaction Hamiltonian has been obtained. We
also gave the numerical analysis of the dynamics based on the full
Hamiltonian within the RWA. In Sec.$\,$V, we give conclusions and point out
some potential applications.

\section{Circuit QED with qutrit}

Here, we show how the multi-atom interactions via the exchange of virtual
photons can be realized in a circuit-QED architecture, where the atoms
strongly coupled to a cavity mode or multiple cavity modes, as shown in Fig.$\,$%
1. In this work, the atoms are treated as cyclic three-level systems,
called qutrit. However, as we discussed later, the third energy levels
of qutrits are actually always empty during the whole process due to
the fact that they serve as auxiliary levels assisting the virtual
photon mediated interactions. Therefore, most of our discussions focus on the lowest
two energy levels of the atoms, which can be considered as qubits. In this
sense, qutrit, qubit, and atom represent different names for a same physical
system in our paper. In experiments, these atoms can be realized by using
the artificial atom such as flux qubit \cite{R38} or fluxonium qubit \cite{R39}.

\subsection{Multi-qutrit coupled with one cavity}
\subsubsection{General model}
\begin{figure}[tbp]
\begin{center}
\includegraphics[width=8.0cm,height=6.0cm]{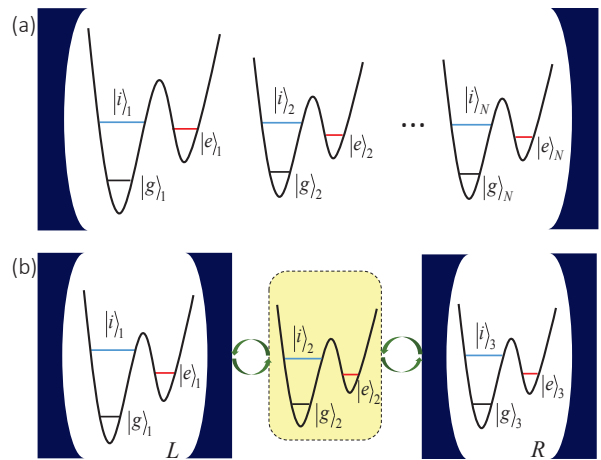}
\end{center}
\caption{(Color online) (a) Schematic of the generic circuit QED system
which consists of $\emph{N}$ qutrits strongly coupled to a cavity mode. The
qutrits are cyclic three-level systems, and the lowest two levels are
treated as qubit states. (b) A schematic of a deformation model for (a) with
three qutrits. The 1st and 3rd qutrit are strongly coupled to
a cavity modes labeled by $L$ and $R$, respectively. The 2nd qutrit
is strongly coupled to the two cavity modes.}
\end{figure}

We first consider a circuit QED system which consists of $N$ atoms
coupled to a cavity mode as depicted in Fig.$\,$1(a). The system can be
described by the Hamiltonian ($\hbar =1$)
\begin{eqnarray}
\begin{aligned}
H=H_{0}+H_{I},
\end{aligned}
\end{eqnarray}
where
\begin{eqnarray}
\begin{aligned}
H_{0}=\omega_{c}a^{\dagger}a+\sum_{q=1}^N\sum_{j=g,e,i}\omega^{(q)}_{j}|j\rangle_{q}\langle j|,
\end{aligned}
\end{eqnarray}
is the Hamiltonian of the cavity mode and the $N$ atoms. $H_{I}$ describes
the atom-cavity interaction
\begin{eqnarray}
\begin{aligned}
H_{I}=\sum_{q=1}^N\sum_{j,k=g,e,i}g^{(q)}_{jk}(a+a^{\dagger})|j\rangle_{q}\langle k|.
\end{aligned}
\end{eqnarray}
Here, $a^{\dagger }$ and $a$ are the creation and annihilation operators for
a cavity mode with frequency $\omega _{c}$, respectively. $\omega _{j}^{(q)}$
is the transition frequency from ground to excited state $|j\rangle _{q}$
for the $q$th atom. $g_{jk}^{(q)}=g_{kj}^{(q)}$ are the coupling strengths
between the $q$th atom and the cavity mode. For easy reference, we set $
\omega _{g}^{(q)}=0$ hereafter.

We assume that our system is operating in the strong coupling regime.
Therefore, we can apply the RWA and drop the counterrotating terms in the
Hamiltonian $H.$ The Hamiltonian can be rewritten as
\begin{eqnarray}
\begin{aligned}
&H=H_{0}+H_{I},
\\&H_{I}=\sum_{q=1}^Na^{\dagger}(g^{(q)}_{ge}|g\rangle_{q}\langle e|+g^{(q)}_{ei}|e\rangle_{q}\langle i|+g^{(q)}_{gi}|g\rangle_{q}\langle i|)+H.c.,
\end{aligned}
\end{eqnarray}
where $H_{0}$ is given in Eq.(2), and the $H.c.$ stands for Hermitian conjugate. Furthermore,
we consider that our system is designed to operate in the dispersive regime,
where the atom-cavity detuning is larger than the coupling strength
between them. Therefore, we have $|\Delta _{jk}^{(q)}|=|(|\omega
_{j}^{(q)}-\omega _{k}^{(q)}|)-\omega _{c}|\gg g_{jk}^{(q)}$. In such a
large detuning regime, coherent conversion of the atom excitation to the
cavity mode can be negligible.

It is worth to notice the difference of Eq.$\,$(4) with that of multiple
two-level systems (qubits) coupled to a cavity. In the latter case, the physics can
be well described by the multi-qubit Tavis-Cummings Hamiltonian under the
usual RWA, which conserves the total excitation number. For the photon-mediated
interactions, the physics picture is that distant qubits can
exchange virtual photons with the cavity bus, leading to nonlocal
interactions between arbitrary two qubits in the cavity. When a pair of
qubits are tuned into resonance, coherent conversion of one qubit excitation
to the other can happen determinatively \cite{R13,R40}. An important feature
is that the total excitation number is conserved in this case.

It is interesting that although the Hamiltonian in Eq.$\,$(4) is also
obtained by using RWA, the Hamiltonian does not conserve the total
excitation number $N_{T}=a^{\dagger }a+\sum_{q=1}^{N}(|e\rangle
_{q}\langle e|+2|i\rangle
_{q}\langle i|)$ \cite{R37}. Therefore, the physical process governed by this
Hamiltonian can violate the conservation law of the excitation number.
Qualitatively, the physics behind the non-conservation of the excitation
number is that the process includes significant contribution from the third
level of the atoms (cyclic qutrits), which allows a large number of virtual
transitions that do not conserve the excitation number. Therefore, Eq.$\,$(4)
provides a approach to coupling three or more atoms via a cavity
bus \cite{R27}. Moreover, when frequency matching condition is satisfied, i.e., $\omega
_{e}^{(1)}=\sum_{q=2}^{N}\omega _{e}^{(q)}$, multiple atoms can jointly
exchange excitation with one single atom. This implies the resonant
transitions between the bare states $|0,e,g,...,g\rangle $ and $
|0,g,e,...,e\rangle $, where for $|0,e,g,...,g\rangle $, the first entry
denotes the cavity state in the Fock representation, while the remain
entries denotes the states of the $N$ atoms.

In order to illustrate the presence and nature of the resonant transition,
we calculated the energy diagram of an example system which consists of
three nondegenerate atoms coupled to a cavity mode. The system is
described by the Hamiltonian in Eq.$\,$(4) with $N=3$. We can obtain the
energy spectrum of the system by numerically solving Schrodinger equation $
H|\Psi _{n}\rangle =E_{n}|\Psi _{n}\rangle $. Shown in Fig.$\,$2 is 6th and
7th eigenenergies as a function of the transition frequency for $|g\rangle
_{1}\leftrightarrow |e\rangle _{1}$, $\omega _{e}^{(1)}/\omega _{0}$, where $
\omega _{0}$ is chosen as a unit of frequency for simplicity. We used $
g_{ge}^{(q)}/\omega _{0}=g_{gi}^{(q)}/\omega _{0}=0.02$, $
g_{ei}^{(q)}/\omega _{0}=0.025$ for all the atoms, $\omega
_{e}^{(2)}/\omega _{0}=0.5$, $\omega _{i}^{(2)}/\omega _{0}=0.9$ for the
$2$nd atom, $\omega _{e}^{(3)}/\omega _{0}=0.55$, $\omega
_{i}^{(3)}/\omega _{0}=1.0$ for the $3$rd atom, $\omega _{i}^{(1)}/\omega
_{0}=1.55$ for the $1$st atom, and $\omega _{c}/\omega _{0}=0.75$ for the
cavity mode \cite{R41}. In this parameter regime, the atom-cavity coupling
strength is much smaller than the atom-cavity detuning. Therefore, the
eigenstates of the Hamiltonian in Eq.$\,$(4) (with $N=3$) can be well
approximated by the bare states \cite{R29,R37}, which are the eigenstates of
the bare Hamiltonian $H_{0}$. As shown in Fig.$\,$(2), we observe that the
two energy levels exhibit an avoided-level crossing, which demonstrates the
resonant coupling between $|0,e,g,g\rangle $ and $|0,g,e,e\rangle $.
Furthermore, we find that far away from the avoided-crossing region one energy level
remains flat as a function of $\omega _{e}^{(1)}/\omega _{0}$ with energy
about $\omega _{e}^{(2)}+\omega _{e}^{(3)}$, while the other growing linearly
with $\omega _{e}^{(1)}/\omega _{0}$. This splitting clearly demonstrates
the hybridization of the states $(|0,e,g,g\rangle \pm |0,g,e,e\rangle )/
\sqrt{2}$ \cite{R31}.

\begin{figure}[tbp]
\begin{center}
\includegraphics[width=8cm,height=6cm]{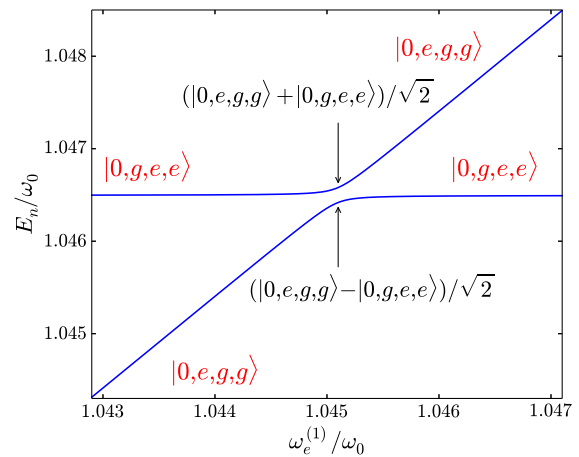}
\end{center}
\caption{(Color online) Calculated energy level diagram of the system which
consists of three nondegenerate qutrits coupled to a cavity mode. Shown here
are $E_{6}$ and $E_{7}$ as functions of $\omega _{e}^{(1)}/\omega _{0}$. An
avoided energy level crossing resulting from the resonant
coupling between $|0,e,g,g\rangle $ and $|0,g,e,e\rangle $ can be observed.
The magnitude of the energy splitting is about $1.8\times 10^{-4}\omega _{0}.$}
\end{figure}

\subsubsection{Effective Hamiltonian}

\begin{figure}[tbp]
\begin{center}
\includegraphics[width=8.0cm,height=6.0cm]{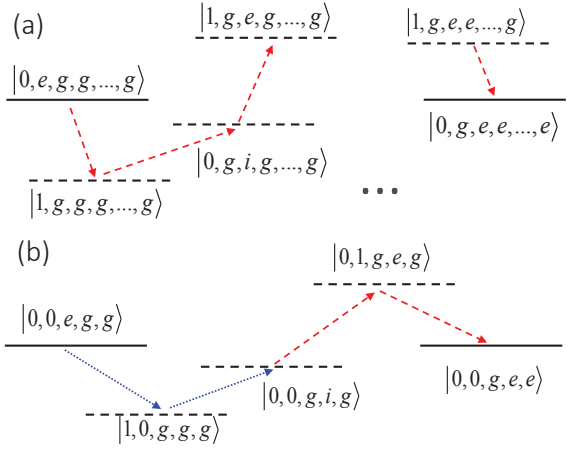}
\end{center}
\caption{(Color online) (a) Sketch of a typical path which contributes in
\emph{2n}th-order perturbation theory to the effective coupling between the
bare states $|0,e,g,...,g\rangle $ and $|0,g,e,...,e\rangle $ for the system
depicted in Fig.$\,$1(a). (b) Sketch of the only path which contributes in
fourth-order perturbation theory to the effective coupling between the bare
states $|0,0,e,g,g\rangle $ and $|0,0,g,e,e\rangle $ for the system depicted
in Fig.$\,$1(b), where the virtual transitions involving the two cavity
modes are represented by dotted blue and dashed red arrows, respectively.
The dashed arrows denote the virtual transitions that do not conserve the
energy, and the dashed black lines represent the intermediate virtual states.}
\end{figure}
Having obtained the general model of our circuit-QED system depicted in Fig.$\,$1(a), we can derive
the effective Hamiltonian which governs various multi-atom interactions by
using the standard perturbation theory \cite{R31,R32}.

We start from the full Hamiltonian in Eq.$\,$(4) which is composed of an
unperturbed part $H_{0}$ with known eigenvalues $E_{i}$ of eigenstates $
|i\rangle ,$ and a small perturbed part $H_{I}$. Following the approach of
Ref. \cite{R32}, we consider an $n$th-order process, which describes the
resonant transitions between the bare states $|i\rangle $ and $|f\rangle $.
The two states are the eigenstates of the bare Hamiltonian $H_{0}$, and have
the same eigenvalues $E_{i}=E_{f}$. We can write the effective interaction
Hamiltonian for this process
\begin{eqnarray}
\begin{aligned}
H_{I}^{eff}=\lambda|f\rangle\langle i|+H.c.,
\end{aligned}
\end{eqnarray}
where $\lambda $ is the effective coupling strength. According to the
standard perturbation theory, the magnitude of the effective coupling
strength can be written as \cite{R32}
\begin{eqnarray}
\begin{aligned}
\lambda=\sum_{j_{1},j_{2},...,j_{n-1}}\frac{V_{fj_{n-1}}...V_{j_{2}j_{1}}V_{j_{1}i}}
{(E_{i}-E_{j_{1}})(E_{i}-E_{j_{2}})...(E_{i}-E_{j_{n-1}})},
\end{aligned}
\end{eqnarray}
where $E_{j_{k}}$ represents the energy of the bare state $|j_{k}\rangle $,
while $V_{j_{k}j_{k+1}}=\langle j_{k}|H_{I}|j_{k+1}\rangle $. The sum goes
over all of the virtual transition steps which forms a transition path
connecting the initial state $|i\rangle $ to the final state $|f\rangle $.

Now, we consider the system introduced earlier in
Sec.$\,$II(A) for the case of $N=n+1$ atoms (labeled by $q$th atom
with $q=1,2,...,n+1$), which is described by the Hamiltonian in
Eq.$\,$(4) with $N=n+1$. The system is initially prepared in state
$|0,e,g,...,g\rangle $. When the frequency matching condition is
satisfied $\omega_{e}^{(1)}\approx \sum_{q=2}^{n+1}\omega _{e}^{(q)}$,
the $n$ atoms ($q$th atom, $q=2,..,n+1$) can conspire to jointly exchange
excitation with the $1$st atom, which implies a resonant interaction
among the multiple atoms via the exchange of virtual photons \cite{R27}.
This process is enabled by the resonant transitions between the bare
states $|0,e,g,...,g\rangle $ and $|0,g,e,...,e\rangle $, with an
example shown in Fig.$\,$(2). By using the $2n$th-order perturbation
theory, and under the frequency matching condition, the system can be
described by the effective Hamiltonian (in the interaction
picture) \cite{R32,R42}
\begin{eqnarray}
\begin{aligned}
H^{I}_{eff}=\lambda_{eff}\sigma_{1}^{-}\sigma_{2}^{+}...\sigma_{n+1}^{+}+H.c.,
\end{aligned}
\end{eqnarray}
where $\sigma _{q}^{\pm }$ are the ladder operators acting on the lowest two levels
$|g\rangle _{q}$ and $|e\rangle _{q}$ of the $q$th atom. $\lambda _{eff}$ is
the effective coupling strength, and its magnitude can be calculated by
using the $2n$th-order perturbation theory given in Eq.$\,$(6).

To illustrate the mechanism of the above process, we present an example path
describing the contribution in $2n$th-order perturbation theory to the
effective coupling between the bare states $|0,e,g,...,g\rangle $ and $
|0,g,e,...,e\rangle $. As depicted in Fig.$\,$3(a), the transition $
|0,e,g,...,g\rangle \longrightarrow |0,g,e,...,e\rangle $ is connected by $2n
$ intermediate virtual transitions, which do not conserve the energy.
Furthermore, except for the $1$st and $(n+1)$th atoms, each atom serves as
a $\Lambda$-type three-level system and gives such a transition path $
|1,g,...,g,...\rangle \longrightarrow |0,g,...,i,...\rangle \longrightarrow
|1,g,...,e,...\rangle ,$ contributing to the effective coupling. After
running through the position of the $n$ atoms ($q$th atom, $q=2,...,n+1$) in every
possible permutation, the final state $|0,g,e,...,e\rangle $ and the initial
state $|0,e,g,...,g\rangle $ are connected via $n\,!$ different transition
paths.

It is worth to mention that: (i) Since our system is initially prepared in
state $|0,e,g,...,g\rangle $, and satisfies the frequency matching
condition, we obtain the resonant effective Hamiltonian under the RWA by
neglecting all the fastvarying terms, and keeping only all terms that are
time independent \cite{R42}. (ii) In our derivation of the effective
interaction Hamiltonian, we have eliminated the third level of the atoms and the
degrees of the freedom of the cavity mode, which are never populated but
nevertheless cause a renormalization and modification of the effective
Hamiltonian. Therefore, we give the effective Hamiltonian in the interaction
picture in Eq.$\,$(7) with respect to the renormalization form of the bare
Hamiltonian. Following the recent work of Guanyu Zhu \textit{et al.} \cite
{R37}, the renormalization form of the bare Hamiltonian $H_{0}$ in the
second-order perturbation theory can be given as
\begin{eqnarray}
\begin{aligned}
H_{0}^{eff}=&\omega_{c}a^{\dagger}a+\sum_{q=1}^{n+1}\sum_{j=g,e,i}
(\omega^{(q)}_{j}+\eta^{(q)}_{j})|j\rangle_{q}\langle j|
\\&+\sum_{q=1}^{n+1}\sum_{j=g,e,i}\xi^{(q)}_{j}a^{\dagger}a|j\rangle_{q}\langle j|,
\end{aligned}
\end{eqnarray}
where the $\xi _{j}^{(q)}$ denotes the ac-Stark type dispersive shifts for
the cavity mode and the $\eta _{j}^{(q)}$ denotes the Lamb type level shift
for the $q$th atom. These coefficients $\xi _{j}^{(q)}$ and $\eta _{j}^{(q)}$
can be determined by using the second-order perturbation theory \cite{R37}.
Certainly, one can further use higher-order perturbation theory to calculate
higher-order correction \cite{R37}.

\subsection{Multi-qutrit coupled through multiple cavities}

From the transition path shown in Fig.$\,$3(a), we find that with the
exception of the atoms involving in the first and the final intermediate
virtual transitions, each atom serves as an $\Lambda -$type three-level
system and contributes such a transition path $|1,g,...,g,...\rangle
\longrightarrow |0,g,...,i,...\rangle \longrightarrow |1,g,...,e,...\rangle $
to the effective coupling. This implies that the one cavity mediated
multi-atom coupling scheme introduced in Sec.$\,$I(A) can be generalized to
multi-cavity case.

As a simple example, we consider a circuit-QED setup shown in Fig.$\,$1(b),
where the two cavities are coupled to a common atom and each cavity hosts
an atom. As shown in Fig.$\,$3(b), we present the transition path in
fourth-order perturbation theory leading to the effective coupling between
the bare states $|0,0,e,g,g\rangle $ and $|0,0,g,e,e\rangle $, where $
|0,0,e,g,g\rangle $ labels the states of the two cavity modes and three
atoms. The path includes four virtual transitions that do not conserve the
energy. The virtual transitions involving the two cavity modes are
represented by dotted blue and dashed red arrows, respectively. Compared
with the one cavity case, for which there are $2$\thinspace $!$ path connecting
the two bare states $|0,e,g,g\rangle $ and $|0,g,e,e\rangle $, there is only
one path which contributes to the effective coupling.

Following the same procedure as in Sec.$\,$I(A), we now turn to
present the quantitatively derivation for the effective coupling among three
atoms mediated by the two cavities. By using RWA, the full system can be
described by the Hamiltonian ($\hbar =1$)
\begin{eqnarray}
\begin{aligned}
H=H_{0}+H_{I},
\end{aligned}
\end{eqnarray}
where
\begin{eqnarray}
\begin{aligned}
H_{0}=\sum_{s=L,R}\omega_{s}a^{\dagger}_{s}a_{s}+\sum_{q=1}^3\sum_{j=g,e,i}\omega^{(q)}_{j}|j\rangle_{q}\langle j|,
\end{aligned}
\end{eqnarray}
describes the Hamiltonian of the two cavities ($s=L,R$), and the three atoms
$(q=1,2,3)$, respectively. $H_{I}$ describes the atom-cavity interaction,
\begin{eqnarray}
\begin{aligned}
H_{I}&=a^{\dagger}_{L}(g^{(1)}_{ge}|g\rangle_{1}\langle e|+g^{(1)}_{ei}|e\rangle_{1}\langle i|+g^{(1)}_{gi}|g\rangle_{1}\langle i|)\\&+\sum_{s=L,R}a^{\dagger}_{s}(g^{(2)}_{ge}|g\rangle_{2}\langle e|+g^{(2)}_{ei}|e\rangle_{2}\langle i|+g^{(2)}_{gi}|g\rangle_{2}\langle i|)\\&+a^{\dagger}_{R}(g^{(3)}_{ge}|g\rangle_{3}\langle e|+g^{(3)}_{ei}|e\rangle_{3}\langle i|+g^{(3)}_{gi}|g\rangle_{3}\langle i|)+H.c.
\end{aligned}
\end{eqnarray}
Here, $a_{s}^{\dagger }$ and $a_{s}$ are the creation and annihilation
operators for the cavity (s) with frequency $\omega _{s}$, respectively. $
\omega _{j}^{(q)}$ is the transition frequency for the $q$th atom from
ground to excited state $|j\rangle _{q}$, and $g_{jk}^{(q)}=g_{kj}^{(q)}$ denote the
atom-cavity coupling strengths for the $q$th atom. For simplicity, we have
assumed that the $2$nd atom is coupled to the two cavity modes ($L,R$)
with the same coupling strength. For easy reference, we set $\omega
_{g}^{(q)}=0$ hereafter.

We consider that the system operates in the dispersive regime, and satisfies
the frequency matching condition $\omega _{e}^{(1)}\approx \omega
_{e}^{(2)}+\omega _{e}^{(3)}$. Following the derivation in Sec.$\,$I(A), we
can write the effective Hamiltonian in the interaction picture
\begin{eqnarray}
\begin{aligned}
H^{I}_{eff}=\scalebox{1.2}{$\chi$}_{2}^{(3)}\sigma_{1}^{-}\sigma_{2}^{+}\sigma_{3}^{+}+H.c.,
\end{aligned}
\end{eqnarray}
where $\sigma _{q}^{\pm }$ are the ladder operators acting on the lowest two
levels $|g\rangle _{q}$ and $|e\rangle _{q}$ of the $q$th atom, and $
\scalebox{1.2}{$\chi$}_{2}^{(3)}$ is the effective coupling strength between
the three atoms. The effective coupling strength is given as
\begin{eqnarray}
\begin{aligned}
\scalebox{1.2}{$\chi$}_{2}^{(3)}=\frac{g_{ge}^{(1)}g_{gi}^{(2)}g_{ei}^{(2)}g_{ge}^{(3)}}
{(\omega^{(1)}_{e}-\omega_{L})(\omega^{(1)}_{e}-\omega^{(2)}_{i})
(\omega^{(1)}_{e}-\omega^{(2)}_{e}-\omega_{R})}.
\end{aligned}
\end{eqnarray}

In principle, the circuit-QED architecture depicted in Fig.$\,$1(b) can also
be generalized to more complex setup, e.g., a two-cavity system where
cavities are coupled to a common atom and each cavity hosts multiple atoms,
and one-dimensional array of the atom-cavity systems depicted in Fig.$\,$
1(b). It allows us to prepare entanglement among atoms located in different
cavities, which are important for large-scale quantum information processing
\cite{R43}. Furthermore, it also allows us to engineer different geometries
for coupling multiple atoms mediated by virtual photons, which can be used
for engineering various lattice systems for quantum simulation \cite{R44}.

\section{application}

In this section, we applied our results derived in Sec.$\,$II to the
concrete case. In particular, we will consider applications in three-atom
case, and discuss the quantum dynamics of the system with experimental
feasible parameters. We would like to mention again that since the third
energy level in principle has null population in the whole process, our
numerical analysis focus on the quantum dynamics of the lowest two energy
levels. Numerical calculation based on the full Hamiltonian under the RWA 
were performed using the PYTHON package QuTiP \cite{R45,R46}. As mentioned
earlier in Sec.$\,$II(A), we did not give analytic expressions for the dispersive
shift of the energy level. However, in the numerical simulation, the effect
of frequency shifts can be canceled out by varying frequencies of the atoms
(in the present work, the frequency shifts are compensated for by modified
frequency for the $1$st atom). Therefore, the frequency matching condition
can be fulfilled.

\subsection{photon-mediated three-atom interaction: one cavity case}

\begin{table}[htbp]
\caption{Parameters for the atom-cavity system described in Sec.$\,$III(A). $
\omega $ is the two-level system transition frequency, $g$ is the atom-cavity
coupling strength, and $\gamma $ is the relaxation rate. For simplicity, we
treat the $2$nd and $3$rd atom as two identical cyclic qutrits.}
\begin{tabular}{p{1.6cm}<{\centering}p{2.0cm}<{\centering}p{2.3cm}<{\centering}p{2.2cm}<{\centering}}
\hline
\hline
Qutrit ($q$=2,3)  & Frequency $\omega/2\pi$ ($GHz$)& Coupling strength $g/2\pi$ ($MHz$)& Relaxation rate $\gamma/2\pi$($MHz$)  \\
\hline
$|g\rangle_{1}\rightarrow|e\rangle_{1} $& 7.966& 150 & 0.01 \\
$|g\rangle_{1}\rightarrow|i\rangle_{1} $& 12.0& 150 & 0.01 \\
$|e\rangle_{1}\rightarrow|i\rangle_{1} $& 4.034& 210 & 0.015 \\
$|g\rangle_{q}\rightarrow|e\rangle_{q} $& 4.0& 150 & 0.01 \\
$|g\rangle_{q}\rightarrow|i\rangle_{q} $& 7.5& 150 & 0.01 \\
$|e\rangle_{q}\rightarrow|i\rangle_{q} $& 3.5& 210 & 0.015 \\
\hline
\hline
\end{tabular}
\end{table}

We consider the system introduced in Sec.II(A) for the case of three
atoms. The system can be described by the Hamiltonian in Eq.$\,$(4)
with $\emph{N}=3$. Furthermore, we assume that $\omega _{e}^{(1)}\approx
\omega _{e}^{(2)}+\omega _{e}^{(3)}$, and the system is initially
prepared in the state $|0,e,g,g\rangle $. The effective coupling between
the three atoms mediated by virtual photon can be described by
\begin{eqnarray}
\begin{aligned}
H^{I}_{eff}=\scalebox{1.2}{$\chi$}_{1}^{(3)}\sigma_{1}^{-}\sigma_{2}^{+}\sigma_{3}^{+}+H.c.,
\end{aligned}
\end{eqnarray}
where $\sigma _{q}^{\pm }$ are the ladder operators acting on the
lowest two levels $|g\rangle _{q}$ and $|e\rangle _{q}$ of the
$q$th atom, and the $\scalebox{1.2}{$\chi$}_{1}^{(3)}$ is the effective
coupling strength between $|0,e,g,g\rangle $ and $|0,g,e,e\rangle $.

In following discussion, we choose system parameters as the typical values
in circuit QED experiments. The frequency of the cavity mode, $\omega
_{c}/2\pi =6.00\,GHz$, with the cavity photon decay rate $\kappa /2\pi
=0.01\,MHz$. The parameter of the three atoms are listed in Table I. We note
that we used two identical qutrits (the $2$nd and $3$rd atom) for the
numerical simulation. This choice is just for the sake of simplicity. System
consisting of three different qutrits can also work. For instance, different
qutrits are used for calculating the energy diagram in Fig.$\,$2. The
presence of resonant transition indicates that we can still have the
interaction.

According to the fourth-order perturbation theory, we can write down the
magnitude of the coupling strength \cite{R32}
\begin{eqnarray}
\begin{aligned}
\scalebox{1.2}{$\chi$}_{1}^{(3)}=\frac{2\,!\,(g_{ge}^{(1)}g_{gi}^{(2)}g_{ei}^{(2)}g_{ge}^{(3)})}
{(\omega^{(1)}_{e}-\omega_{c})(\omega^{(1)}_{e}-\omega^{(2)}_{i})
(\omega^{(1)}_{e}-\omega^{(2)}_{e}-\omega_{c})},
\end{aligned}
\end{eqnarray}
which results $\scalebox{1.2}{$\chi$}_{1}^{(3)}/2\pi \approx 0.760\,MHz$. As
shown in Table$\,$I, we note that since the $2$nd and $3$rd atoms are
treated as two identical cyclic three-level systems, there are $2\,!$ paths
which have equal contributions to the effective coupling between the bare
state $|0,e,g,g\rangle $ and $|0,g,e,e\rangle $.

\begin{figure}[tbp]
\begin{center}
\includegraphics[width=8cm,height=6cm]{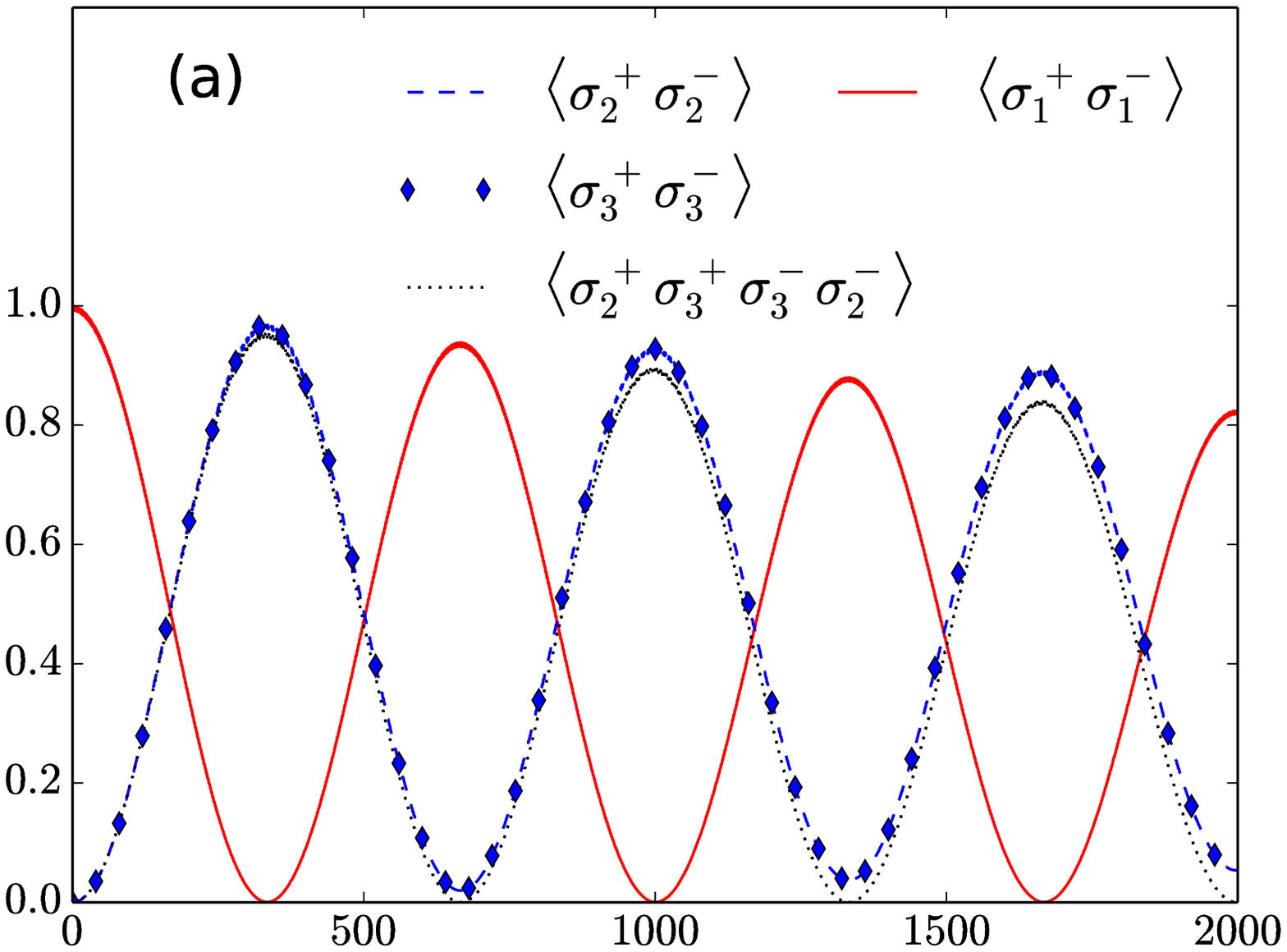}
\includegraphics[width=8cm,height=2cm]{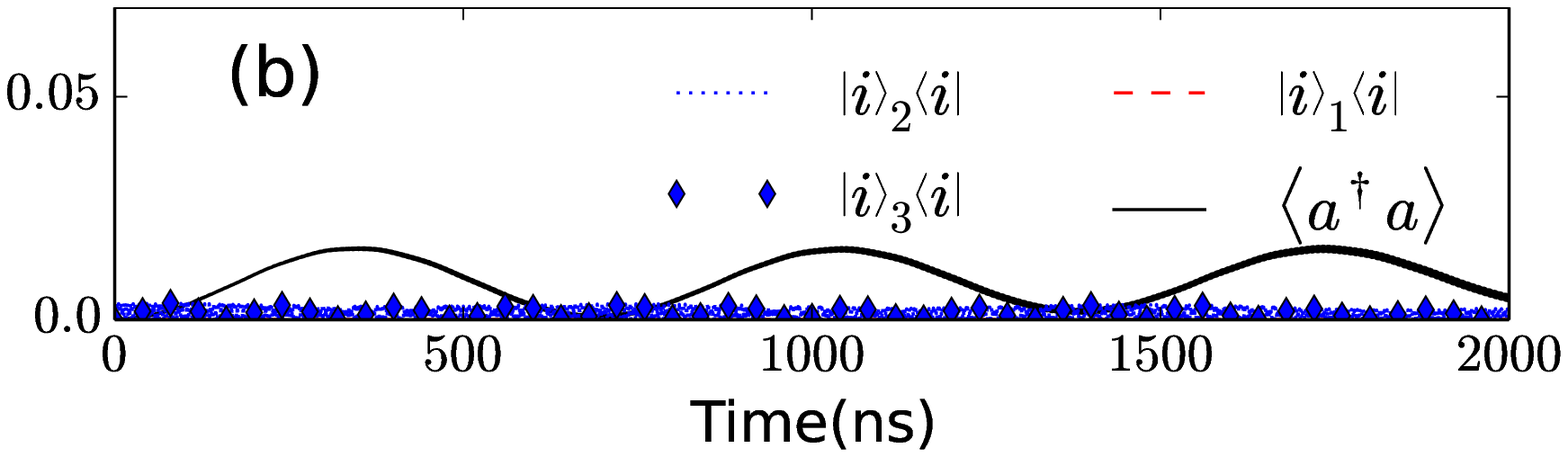}
\end{center}
\caption{(Color online) Numerical simulation of the dynamics under the
influence of dissipation. (a) Temporal evolution of the atom mean
excitation number $\langle\sigma _{q}^{+}\protect\sigma _{q}^{-}\rangle$,
and the equal-time second-order correlation function $
\langle\sigma _{2}^{+}\sigma _{3}^{+}\sigma _{3}^{-}\sigma _{2}^{-}\rangle $
with the system prepared in the state $\left\vert 0,e,g,g\right\rangle $. (b) The
residual population in the third level of the atoms $|i\rangle_{q}\langle i|$
and the cavity mode $\langle a^{\dag}a\rangle$.}
\end{figure}

We numerically simulated the dynamics of the system under the influence of
cavity decay and atom relaxation by using the master equation approach (see
Appendix B). The numerical calculations have been performed based on the
Hamiltonian in Eq.$\,$(4) with $\emph{N}=3$. Shown in Fig.$\,$4(a) is the
time evolution of the atom mean excitation number $\langle \sigma _{q}^{+}\sigma _{q}^{-}\rangle $. It can
be observed from this ordinary oscillation between $|0,e,g,g\rangle $ and $
|0,g,e,e\rangle $ that two atoms can jointly exchange excitation with just
one single atom in a reversible and coherent way. The period of the
oscillation is in good agreement with the value calculated based on the
effective coupling strength $\scalebox{1.2}{$\chi$}_{1}^{(3)}$, i.e., $T=\pi
/\scalebox{1.2}{$\chi$}_{1}^{(3)}\approx 658$ $ns$. In Fig.$\,$4(b), we also
display the population leakage to the third level of the atoms and the bus
cavity. During the time evolution, the population leakage in the third level
of the atoms can be negligible, and the mean photon number is far less than
1.

Figure$\,$4(a) shows the time evolution of the two-atom correlation function $
\langle \sigma _{2}^{+}\sigma _{3}^{+}\sigma _{3}^{-}\sigma _{2}^{-}\rangle $
, which describes the quantum correlation between the emitted photons from
the two atoms into noncavity modes \cite{R31}. We observe that the two-atom
correlation function and the atom mean excitation number $\langle \sigma
_{q}^{+}\sigma _{q}^{-}\rangle $ $(q=2,3)$ almost coincide at early time
\cite{R27}. This is a signature of almost perfect two-atom correlation: if
one atom is excited, the other is also excited \cite{R31}. We also find that
the two-atom correlation function goes almost to zero every time the mean
excitation number of the $1$st atom $\langle \sigma _{1}^{+}\sigma
_{1}^{-}\rangle $ has maximally value. This behavior indicates that the
excitation of the $1$st atom does not convert to a single atom (the $2$nd
atom or the $3$rd atom) but the two atom jointly \cite{R47}. In the Appendix
A, we also consider the four-atom case, and obtain similar results.

This virtual photon-mediated three-atom interaction allows for the
realization of three-atom entanglement. Figure$\,$4(a) implies that when the
system is initialized in state $|e,g,g\rangle $, which labels the three
atoms state, the entangled state $(|e,g,g\rangle +|g,e,e\rangle )/\sqrt{2}$
can be obtained after a time $t=\pi /(4\scalebox{1.2}{$\chi$}_{1}^{(3)})$.
Moreover, along with the single atom operations, one can create the
three-atom GHZ state $(|g,g,g\rangle +|e,e,e\rangle )/\sqrt{2}$.

\subsection{photon-mediated three-atom interaction: two-cavity case}

Here, we give the numerical analysis of the atom-cavity system depicted in
Fig.$\,$1(b). Our aim is to give a numerical confirmation of the theoretical
demonstration of the two-cavity mediated three-atom interaction discussed
in Sec.$\,$II(B). The numerical calculations are performed based on the
Hamiltonian in Eq.$\,$(9). Furthermore, we note that the $1$st and $3$rd atom acted as two-level
systems formed by the lowest two levels of the atoms, while the $2$nd atom
worked as an $\Lambda $-type three-level system in the
transition path leading to the three-atom interaction, as shown in Fig.$\,$
3(b). Therefore, for simplicity, we can treat the $1$st and $3$rd atom as
two two-level systems in the numerical calculations. We would like to mention again that
in the following discussion, $\sigma _{q}^{\pm }$ are the ladder
operators acting on the lowest two levels $|g\rangle _{q}$ and
$|e\rangle _{q}$ of the $q$th atom as we defined in Sec.$\,$II(B).

In Fig.$\,$5, we show the time evolution of the atom mean excitation number $
\langle \sigma _{q}^{+}\sigma _{q}^{-}\rangle $ and the two-atom correlation
function $\langle \sigma _{2}^{+}\sigma _{3}^{+}\sigma _{3}^{-}\sigma
_{2}^{-}\rangle $ under the influence of cavity decay and atom relaxation.
We also display the population leakage in the third level of the $2$nd atom
($|i\rangle _{2}$) and cavity photon population $\langle a_{s}^{\dag
}a_{s}\rangle \,(L,R)$, as shown in Fig.$\,$5(b). For the numerical
simulation, the system parameters are chosen as follows, $\omega _{L}/2\pi
=6.00\,GHz$ and $\omega _{R}/2\pi =6.00\,GHz$ are resonance frequency of the
two cavity modes $(L,R)$, respectively. $\omega _{e}^{(1)}/2\pi \approx
7.945\,GHz$ is the transition frequency for the $1$st atom, and $\omega
_{e}^{(3)}/2\pi =4.00\,GHz$ is for the $3$rd atom. The transition
frequencies of the $2$nd atom are $\omega _{i}^{(2)}/2\pi =7.50\,GHz$ and
$\omega _{e}^{(2)}/2\pi =4.00\,GHz$. $g_{ge}^{(1)}/2\pi =g_{ge}^{(3)}/2\pi
=180\,MHz$, $g_{ge}^{(2)}/2\pi =g_{gi}^{(2)}/2\pi =150\,MHz$, and $
g_{ei}^{(2)}/2\pi =210\,MHz$ are the atom-cavity coupling strengths. The
cavity photon decay rate and the atom relaxation rate are $\kappa _{s}/2\pi
=\gamma _{ge}^{(q)}/2\pi =\gamma _{gi}^{(2)}/2\pi =0.01\,MHz$, and $\gamma
_{ei}^{(2)}/2\pi =0.015\,MHz$, respectively.

Similar to the one cavity case, the two-cavity mediated coherent conversion of
the $1$st atom excitation to the other two atoms (the $2$nd and $3$rd
atoms) can happen determinatively, and the two-atom correlation function
also coincide with the atom mean excitation number $\langle \sigma
_{q}^{+}\sigma _{q}^{-}\rangle $ $(q=2,3)$ at early time, as shown in Fig.$\,
$5(a). We can find that the period of the oscillation which is calculated
based on the effective coupling strength $\scalebox{1.2}{$\chi$}_{2}^{(3)}$,
i.e., $T=\pi /\scalebox{1.2}{$\chi$}_{2}^{(3)}\approx 871$ $ns$, is good
agreement with the result of the numerical simulation of the full dynamics.
Figure$\,$5(b) demonstrates that the population leakage in the third level of
the $2$nd atom can be negligible, and the photon population in the two
cavity modes $(L,R)$ is far less than 1.

\begin{figure}[tbp]
\begin{center}
\includegraphics[width=8cm,height=6cm]{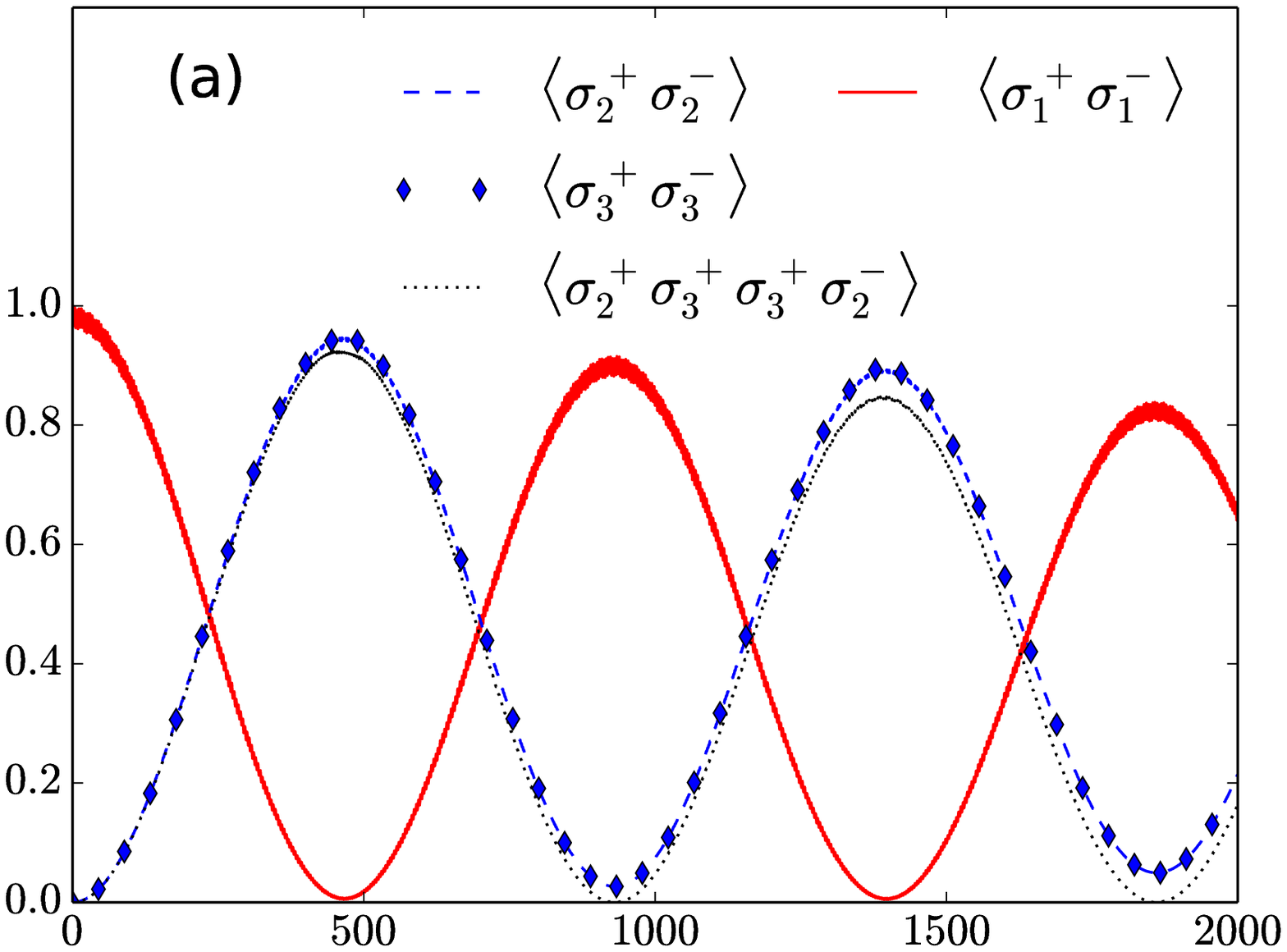}
\includegraphics[width=8cm,height=2cm]{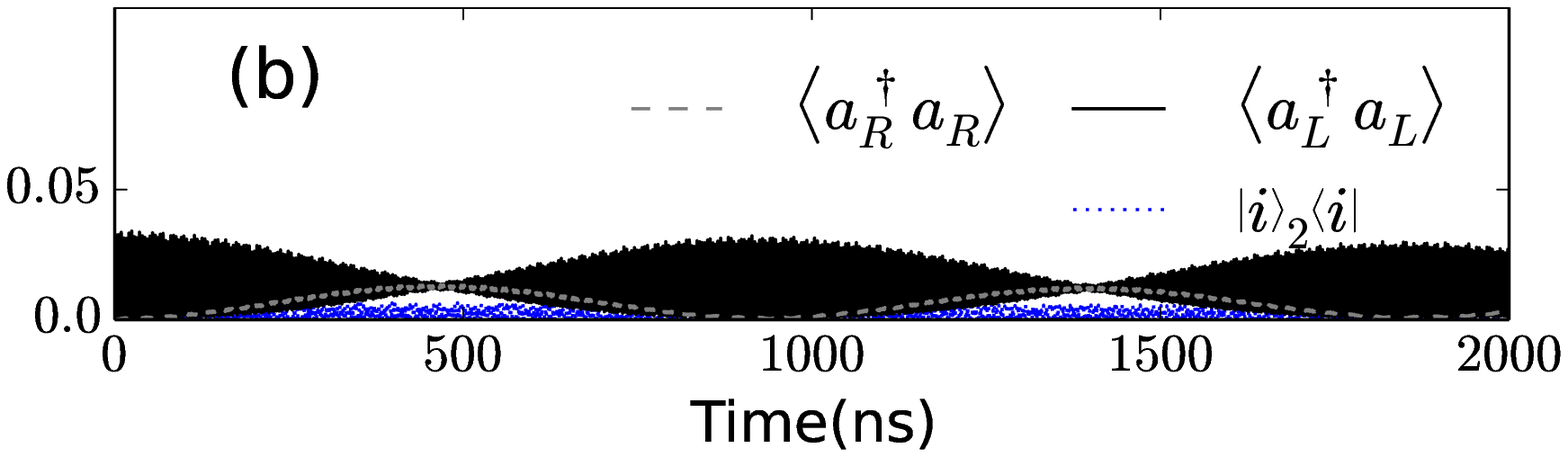}
\end{center}
\caption{(Color online) Numerical simulation of the dynamics under the
influence of dissipation. (a) Temporal evolution of the atom mean excitation
number $\langle \protect\sigma _{q}^{+}\protect\sigma _{q}^{-}\rangle $, and
the equal-time second-order correlation function $\langle\sigma_{2}^{+}
\sigma _{3}^{+}\sigma _{3}^{-}\sigma_{2}^{-}\rangle $ with the system
prepared in the state $|0,0,e,g,g\rangle$. (b) The residual population in the third level of
the $2$nd atom $|i\rangle _{2}\langle i|$ and the two cavity modes $\langle
a_{s}^{\dag }a_{s}\rangle $ $(L,R)$.}
\end{figure}

\section{Conclusion}

In summary, we have studied the resonant exchange interaction among three or
more atoms via the exchange of virtual photons in circuit QED system
consisting of multi atoms strongly coupled to a cavity mode. If the
selection rule of the atom transitions is violated, multiple atoms can
jointly exchange excitation with just one single atom in a reversible and
coherent way. The analytical and numerical results demonstrate that this
excitation number nonconserving process can happen with probability
approaching one. In addition, we show that the two-cavity mediated three-atom exchange
interaction can also be realized in the strong coupling regime.

This process can be exploited for the realization of an efficient atom-atom
entanglement source \cite{R27,R46} and can also be used for the
implementation of novel schemes for the control and manipulation of atom
states, e.g., three-qubit gates \cite{R48}, quantum repetition coding \cite
{R27} needed for error-correction codes. Furthermore, it is also possible to
use the photon-mediated multi-atom interaction to engineer long-distance
entangled state and stabilization of pure many-body states of atoms \cite
{R49}.

In practical, the difficulties one may expect to face with these higher
order processes depend on the limitations inherent with the decoherence
process of the atom (cyclic qutrit). However, with recent experimental progress in circuit
QED, especially, superconducting qubits with a long coherence time have been experimentally
demonstrated \cite{R50,R51,R52,R53,R54}, we estimate that our proposed
architecture with three or four atoms is feasible with currently available
technology.

\acknowledgements This work was partly supported by the NKRDP of China
(Grant No. 2016YFA0301802) and NSFC (Grants No. 11504165, No. 11474152, and No.
61521001).

\appendix

\section{Cavity-photon mediated four-atom interaction}

\begin{table}[htbp]
\caption{Parameters for the atom-cavity system described in Appendix A. $
\omega $ is the two-level system transition frequency, $g$ is the
atom-cavity coupling strength, and $\protect\gamma $ is the relaxation
rate. For simplicity, we treat the $2$nd atom, $3$rd atom,
and $4$th atom as three identical cyclic three-level systems.}
\begin{tabular}{p{1.6cm}<{\centering}p{2.0cm}<{\centering}p{2.3cm}<{\centering}p{2.2cm}<{\centering}}
\hline
\hline
Qutrit ($q$=2,3,4)  & Frequency $\omega/2\pi$ ($GHz$)& Coupling strength $g/2\pi$ ($MHz$)& Relaxation rate $\gamma/2\pi$($MHz$)  \\
\hline
$|g\rangle_{1}\rightarrow|e\rangle_{1} $& 8.9665& 180 & 0.01 \\
$|g\rangle_{1}\rightarrow|i\rangle_{1} $& 21.0& 180 & 0.01 \\
$|e\rangle_{1}\rightarrow|i\rangle_{1} $& 12.0335& 210 & 0.015 \\
$|g\rangle_{q}\rightarrow|e\rangle_{q} $& 3.0& 150 & 0.01 \\
$|g\rangle_{q}\rightarrow|i\rangle_{q} $& 7.0& 150 & 0.01 \\
$|e\rangle_{q}\rightarrow|i\rangle_{q} $& 4.0& 200 & 0.015 \\
\hline
\hline
\end{tabular}
\end{table}

In this appendix, we provide a discussion on the photon-mediated four-atom
interaction. We consider the system introduced in Sec.$\,$II(A) of the main text
for the case of four atoms. The system can be described by the Hamiltonian
in Eq.$\,$(4) with $N=4$. For easy reference, we set $\omega _{g}^{(q)}=0$
hereafter. Furthermore, we consider that the system is initialized in state $
|0,e,g,g,g\rangle $, and satisfies the frequency matching condition $\omega
_{e}^{(1)}\approx \omega _{e}^{(2)}+\omega _{e}^{(3)}+\omega _{e}^{(4)}$.
The parameter of the four atom are listed in Table$\,$II, the resonance
frequency of the cavity mode is $\omega _{c}/2\pi =6.00\,GHz$, and the
cavity photon decay rate is $\kappa /2\pi =0.01\,MHz$.

Under the frequency matching condition, and by using the sixth-order
perturbation theory, the system can be described by the effective
Hamiltonian (in the interaction picture)
\begin{eqnarray}
\begin{aligned}
H^{I}_{eff}=\scalebox{1.2}{$\chi$}_{1}^{(4)}\sigma_{1}^{-}\sigma_{2}^{+}\sigma_{3}^{+}\sigma_{4}^{+}+H.c.,
\end{aligned}
\end{eqnarray}
where $\sigma _{q}^{\pm }$ ($q=1,2,3,4$) are the ladder operators acting on the lowest
two levels $|g\rangle _{q}$ and $|e\rangle _{q}$ of the $q$th atom, and
the $\scalebox{1.2}{$\chi$}_{1}^{(4)}$ is the effective coupling strength
between $|0,g,e,e,e\rangle $ and $|0,e,g,g,g\rangle $. The magnitude of the
effective coupling strength can be written as
\begin{widetext}
\begin{eqnarray}
\begin{aligned}
\scalebox{1.2}{$\chi$}_{1}^{(4)}=\frac{3\,!\,(g_{ge}^{(1)}g_{gi}^{(2)}g_{ei}^{(2)}g_{gi}^{(3)}g_{ei}^{(3)}g_{ge}^{(4)})}
{(\omega^{(1)}_{e}-\omega_{c})(\omega^{(1)}_{e}-\omega^{(2)}_{i})(\omega^{(1)}_{e}-\omega^{(2)}_{e}-\omega_{c})
(\omega^{(1)}_{e}-\omega^{(3)}_{i}-\omega^{(2)}_{e})(\omega^{(1)}_{e}-\omega^{(3)}_{e}-\omega^{(2)}_{e}-\omega_{c})}.
\end{aligned}
\end{eqnarray}
\end{widetext}
After calculation with the above-mentioned system parameters,
the effective coupling strength is obtained as $\scalebox{1.2}{$\chi$}
_{1}^{(4)}/2\pi \approx 0.238\,MHz$. As shown in Table$\,$II, we note that
since the three atoms ($2$nd, $3$rd, and $4$th atom) are treated as
three identical cyclic qutrits, there are $3\,!$ paths which have equal
contributions to the effective coupling between the bare state $
|0,e,g,g,g\rangle $ and $|0,g,e,e,e\rangle $.

As shown in Fig.$\,$6(a), it can be observed from the oscillation of the
atom mean excitation number that three atoms can conspire to jointly
exchange excitation with one single atom, and the period is good agreement
with calculation given by the perturbation theory. The three-atom
correlation function $\langle \sigma _{2}^{+}\sigma _{3}^{+}\sigma
_{4}^{+}\sigma _{4}^{-}\sigma _{3}^{-}\sigma _{2}^{-}\rangle $ and atom
mean excitation number $\langle \sigma _{2}^{+}\sigma _{2}^{-}\rangle$
are almost coincident at early time \cite{R27}. Figure$\,$6(b) shows that
the residual population in the third level of the four atoms can be
negligible, and the population in the cavity bus is far less than 1.

As we mentioned in Sec.$\,$I of the main text, there exist one other type of
effective interactions between the four atoms in the fourth-order
perturbation terms, namely, the four-spin ring exchange interaction $\sigma
_{1}^{+}\sigma _{2}^{+}\sigma _{3}^{-}\sigma _{4}^{-}$ \cite{R26,R27,R55}.
Noted that the total excitation number is conserved in this case. Therefore,
the four-spin ring exchange interaction can be simply captured by the
four-atom Tavis-Cummings Hamiltonian within the RWA.

\begin{figure}[htp]
\begin{center}
\includegraphics[width=8cm,height=6.0cm]{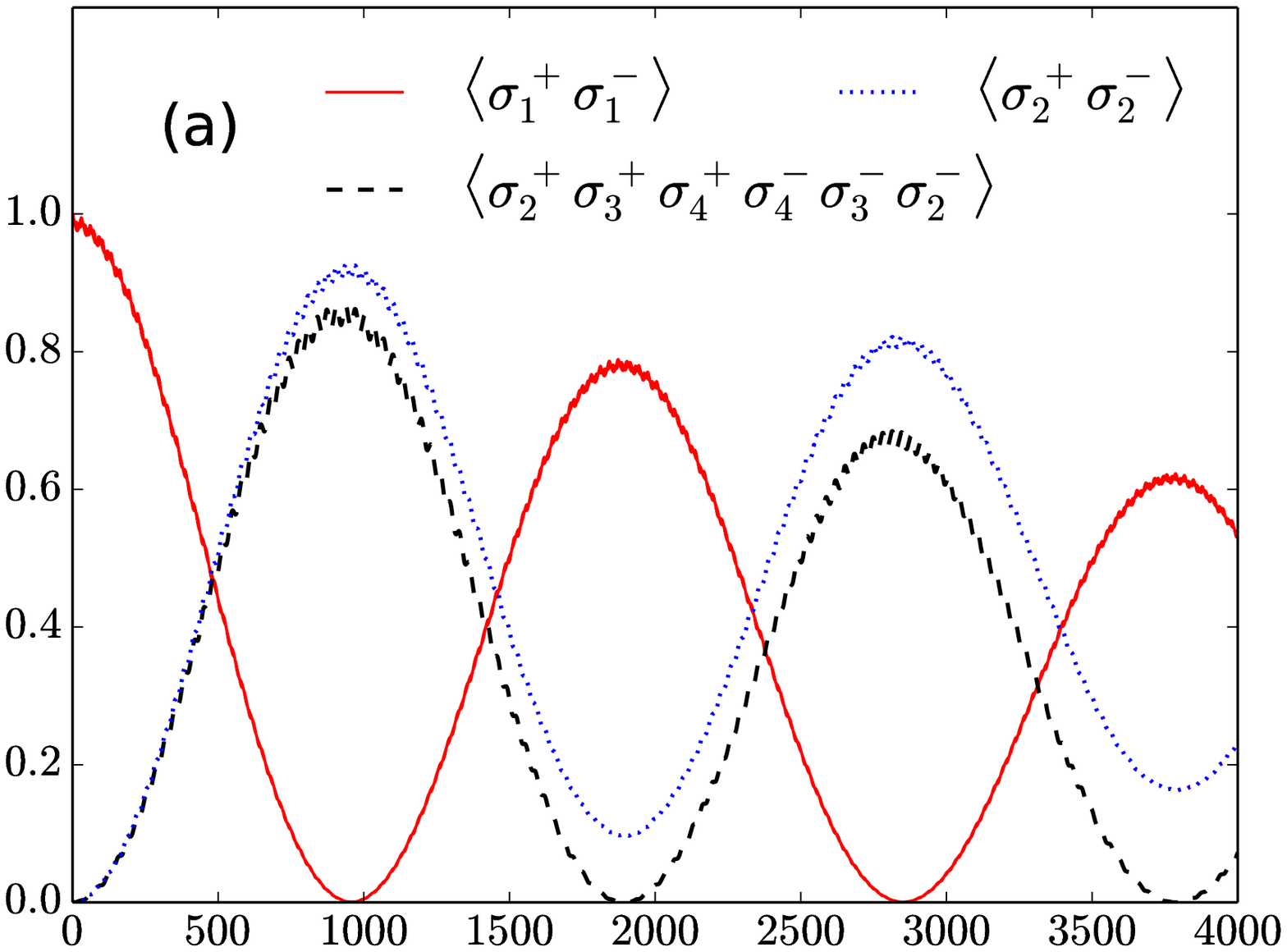}
\includegraphics[width=8cm,height=2.0cm]{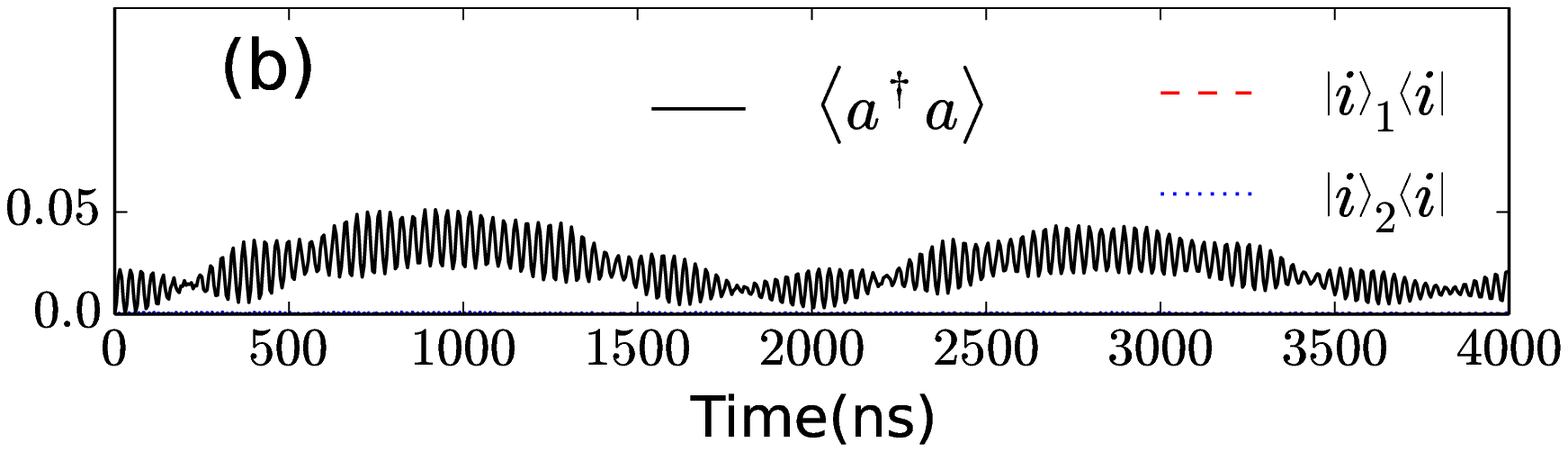}
\end{center}
\caption{(Color online) Numerical simulation of the dynamics under the
influence of dissipation. (a) Temporal evolution of the atom mean excitation
number $\langle \protect\sigma _{q}^{+}\sigma _{q}^{-}\rangle $, and
the three-atom correlation function $\langle\sigma _{2}^{+}\sigma _{3}^{+}
\sigma _{4}^{+}\sigma _{4}^{-}\sigma_{3}^{-}\sigma _{2}^{-}\rangle $ with
the system prepared in the state $|0,e,g,g,g\rangle $. (b) the residual population in
the third level of the atoms $|i\rangle _{q}\langle i|$ and the cavity
mode $\langle a^{\dag }a\rangle $.}
\end{figure}

\section{the master equation approach}

The influence of cavity decay and atom relaxation on the process can be
studied by the master equation approach. By including cavity decay and atom
relaxation terms, we can write the master equation:
\begin{eqnarray}
\begin{aligned}
\frac{d\rho}{dt}=&-i[H,\rho]+\sum_{s}\kappa_{s}\mathcal{L}[a_{s}]+\sum_{q}(\gamma^{(q)}_{ge}\mathcal{L}[|g\rangle _{q}\langle e|]\\&+\gamma^{(q)}_{ei}\mathcal{L}[|e\rangle _{q}\langle i|]+\gamma^{(q)}_{gi}\mathcal{L}[|g\rangle _{q}\langle i|]).
\end{aligned}
\end{eqnarray}
Above, $\rho$ is the reduced density matrix of the system, $H$ is the
Hamiltonian of the system, $\mathcal{L}[O]=O\rho O^{\dagger }-O^{\dagger }
O\rho/2-\rho O^{\dagger }O/2$, $\kappa_{s} $ and $\gamma _{jk}^{(q)}$ denote
the photon decay rate of the cavity mode ($s$) and the relaxation rate of the $(|j\rangle
_{q},|k\rangle _{q} )$ two level systems, respectively.


\begin{thebibliography}{99}


\bibitem{R1} C. Cohen-Tannoudji, J. Dupont-Roc, and G. Grynberg, Atom-Photon Interactions: Basic Processes and Applications (Wiley, New York, 1992).
\bibitem{R2} T. D. Ladd, F. Jelezko, R. Laflamme, Y. Nakamura, C. Monroe, and J. L. O¡¯Brien, Nature (London) \textbf{464}, 45 (2010).
\bibitem{R3} H. J. Kimble, Nature \textbf{453}, 1023 (2008).
\bibitem{R4} A. Blais, J. Gambetta, A. Wallraff, D. I. Schuster, S. M. Girvin, M. H. Devoret, and R. J. Schoelkopf, Phys. Rev. A \textbf{75}, 032329 (2007).
\bibitem{R5} J. Q. You and Franco Nori, Phys. Rev. B \textbf{68}, 064509 (2003).
\bibitem{R6} J. Q. You and F. Nori, Nature (London) \textbf{474}, 589 (2011).
\bibitem{R7} M. H. Devoret and R. J. Schoelkopf, Science \textbf{339}, 1169 (2013).
\bibitem{R8} A. S$\o$rensen and K. M$\o$lmer, Phys. Rev. Lett. \textbf{82}, 1971 (1999).
\bibitem{R9} A. Imamoglu, D. D. Awschalom, G. Burkard, D. P. DiVincenzo, D. Loss, M. Sherwin, and A. Small, Phys. Rev. Lett. \textbf{83}, 4204 (1999).
\bibitem{R10} S.-B. Zheng and G.-C. Guo, Phys. Rev. Lett. \textbf{85}, 2392 (2000).
\bibitem{R11} A. Blais, R.-S. Huang, A. Wallraff, S. M. Girvin, and R. J. Schoelkopf, Phys. Rev. A \textbf{69}, 062320 (2004).
\bibitem{R12} L. DiCarlo, J. M. Chow, J. M. Gambetta, L. S. Bishop, B. R. Johnson, D. I. Schuster, J. Majer, A. Blais, L. Frunzio, S. M. Girvin, and R. J. Schoelkopf, Nature (London) \textbf{460}, 240 (2009).
\bibitem{R13} J. Majer, J. M. Chow, J. M. Gambetta, J. Koch, B. R. Johnson,
    J. A. Schreier, L. Frunzio, D. I. Schuster, A. A. Houck, A. Wallraff, A.
    Blais, M. H. Devoret, S. M. Girvin, and R. J. Schoelkopf, Nature (London)
    \textbf{449}, 443 (2007).
\bibitem{R14} L. DiCarlo, M. D. Reed, L. Sun, B. R. Johnson, J. M. Chow, J. M. Gambetta, L. Frunzio, S. M. Girvin, M. H. Devoret, and R. J. Schoelkopf, Nature (London) \textbf{467}, 574 (2010).
\bibitem{R15} M. Neeley, R. C. Bialczak, M. Lenander, E. Lucero, M. Mariantoni, A. D. O'Connell, D. Sank, H. Wang, M. Weides, J. Wenner, Y. Yin, T. Yamamoto, A. N. Cleland, and J. M. Martinis, Nature (London) \textbf{467}, 570 (2010).
\bibitem{R16} C. H. Bennett and D. P. DiVincenzo, Nature (London) \textbf{404}, 247 (2000).
\bibitem{R17} A. M. Chen, S. Y. Cho, and M. D. Kim, Phys. Rev. A \textbf{85}, 032326 (2012).
\bibitem{R18} E. Zahedinejad, J. Ghosh, and B. C. Sanders, Phys. Rev. Lett. \textbf{114}, 200502 (2015).
\bibitem{R19} E. Barnes, C. Arenz, A. Pitchford, and S. E. Economou. arXiv:1612.09384 (2016).
\bibitem{R20} H. Weimer, M. M\"{u}ller, I. Lesanovsky, P. Zoller, and H. P. B\"{u}chler, Nat. Phys. \textbf{6}, 382 (2010).
\bibitem{R21} A. Mezzacapo, L. Lamata, S. Filipp, and E. Solano, Phys. Rev. Lett. \textbf{113}, 050501 (2014).
\bibitem{R22} D. Becker, T. Tanamoto, A. Hutter, F. L. Pedrocchi, and D. Loss, Phys. Rev. A \textbf{87}, 042340 (2013).
\bibitem{R23} T. Tanamoto, V. M. Stojanovi\'{c}, C. Bruder, and D. Becker, Phys. Rev. A \textbf{87}, 052305 (2013).
\bibitem{R24} E. T. Jaynes and F. W. Cummings, Proc. IEEE \textbf{51}, 89
    (1963).
\bibitem{R25} M. Tavis and F. W. Cummings, Phys. Rev. \textbf{170}, 379 (1968).
\bibitem{R26} G. Zhu, M. Hafezi, and T. Grover, Phys. Rev. A, \textbf{94}, 062329 (2016).
\bibitem{R27} R. Stassi, V. Macr\`{\i}, A. F. Kockum, O. Di Stefano, A. Miranowicz, S. Savasta, and F. Nori, arXiv:1702.00660 (2017).
\bibitem{R28} J. Bourassa, J. M. Gambetta, A. A. Abdumalikov, O. Astafiev,
    Y. Nakamura, and A. Blais, Phys. Rev. A \textbf{80}, 032109 (2009).
\bibitem{R29} K. K. W. Ma and C. K. Law, Phys. Rev. A \textbf{92}, 023842 (2015).
\bibitem{R30} L. Garziano, R. Stassi, V. Macr\`{\i}, A. F. Kockum, S. Savasta, and F. Nori, Phys. Rev. A \textbf{92}, 063830 (2015).
\bibitem{R31} L. Garziano, V. Macr\`{\i}, R. Stassi, O. Di Stefano, F. Nori, and S. Savasta, Phys. Rev. Lett. \textbf{117(4)}, 043601 (2016).
\bibitem{R32} A. F. Kockum, A. Miranowicz, V. Macr\`{\i}, S. Savasta, and F. Nori, arXiv:1701.05038 (2017).
\bibitem{R33} A. F. Kockum, V. Macr\`{\i}, L. Garziano, S. Savasta, and F. Nori, arXiv:1701.07973 (2017).

\bibitem{R34} T. Niemczyk, F. Deppe, H. Huebl, E. P. Menzel, F. Hocke, M. J.
    Schwarz, J. J. Garc\'{\i}a-Ripoll, D. Zueco, T. H\"{u}mmer, E. Solano, A.
    Marx, and R. Gross, Nature Phys. \textbf{6}, 772 (2010).
\bibitem{R35} P. Forn-D\'{\i}az, J. Lisenfeld, D. Marcos, J. J. Garc\'{\i}%
    a-Ripoll, E. Solano, C. J. P. M. Harmans, and J. E. Mooij, Phys. Rev. Lett.
    \textbf{105}, 237001 (2010).
\bibitem{R36} A. Fedorov, A. K. Feofanov, P. Macha, P. Forn-D\'{i}az, C. J. P. M. Harmans, and J. E. Mooij, Phys. Rev. Lett. \textbf{105}, 060503 (2010).
\bibitem{R37} G. Zhu, D. G. Ferguson, V. E. Manucharyan, and J. Koch, Phys. Rev. B \textbf{87}, 024510 (2013).
\bibitem{R38} Y.-X. Liu, J. Q. You, L. F. Wei, C. P. Sun, and F. Nori, Phys. Rev. Lett. \textbf{95}, 087001 (2005).
\bibitem{R39} V. E. Manucharyan, J. Koch, L. I. Glazman, and M. H. Devoret, Science \textbf{326}, 113 (2009).
\bibitem{R40} M. R. Delbecq, L. E. Bruhat, J. J. Viennot, S. Datta, A. Cottet, and T. Kontos, Nat. Commun. \textbf{4}, 1400 (2013).
\bibitem{R41} T. Yamamoto, K. Inomata, K. Koshino, P.-M. Billangeon, Y. Nakamura, and J. S. Tsai, New J. Phys. \textbf{16}, 015017 (2014).
\bibitem{R42} C. K. Law, Phys. Rev. A \textbf{49}, 433 (1994).
\bibitem{R43} C. P. Yang, Q. P. Su, S. B. Zheng, and F. Nori, New J. Phys. \textbf{18}, 013025 (2016).
\bibitem{R44} D. I. Tsomokos, S. Ashhab, and F. Nori, Phys. Rev. A \textbf{82}, 052311 (2010).
\bibitem{R45} J. R. Johansson, P. D. Nation, and F. Nori, Compu. Phys.
    Commun. \textbf{183}, 1760 (2012).
\bibitem{R46} J. R. Johansson, P. D. Nation, and F. Nori, Compu. Phys. Comm.
    \textbf{184}, 1234 (2013).
\bibitem{R47} L. Garziano, R. Stassi, V. Macr\`{\i}, A. F. Kockum, S.
    Savasta, and F. Nori, Phys. Rev. A \textbf{92}, 063830 (2015).
\bibitem{R48} C. T. Rigetti, Ph.D. thesis, Yale University, New Haven, Connecticut, 2009.
\bibitem{R49} C. Aron, M. Kulkarni, and H. E. T\"{u}reci, Phys. Rev. X \textbf{6}, 011032 (2016)
\bibitem{R50} F. Yan, S. Gustavsson, A. Kamal, J. Birenbaum, A. P. Sears, D. Hover, T. J. Gudmundsen,
              D. Rosenberg, G. Samach, S. Weber, J. L. Yoder, T. P. Orlando, J. Clarke, A. J. Kerman,
              and W. D. Oliver, Nat.Commun. \textbf{7}, 12964 (2016).
\bibitem{R51} J.-L. Orgiazzi, C. Deng, D. Layden, R. Marchildon, F. Kitapli, F. Shen, M. Bal, F. Ong, and A. Lupascu, Phys. Rev. B \textbf{93}, 104518 (2016).
\bibitem{R52} M. Stern, G. Catelani, Y. Kubo, C. Grezes, A. Bienfait, D. Vion, D. Esteve, and P. Bertet, Phys. Rev. Lett. \textbf{113}, 123601 (2014).
\bibitem{R53} I. M. Pop, K. Geerlings, G. Catelani, R. J. Schoelkopf, L. I. Glazman, and M. H. Devoret, Nature (London) \textbf{508}, 369 (2014).
\bibitem{R54} U. Vool, I. M. Pop, K. Sliwa, B. Abdo, C. Wang, T. Brecht, Y. Y. Gao, S. Shankar, M. Hatridge, G. Catelani, M. Mirrahimi, L. Frunzio, R. J. Schoelkopf, L. I. Glazman, and M. H. Devoret, Phys. Rev. Lett. \textbf{113}, 247001 (2014)
\bibitem{R55} M. Sameti, A. Potocnik, D. E. Browne, A. Wallraff, and M. J. Hartmann, Phys. Rev. A \textbf{95}, 042330 (2017).

\end{thebibliography}
\end{document}